\documentclass[english,apm, reprint, twocolumn, superscriptaddress,showpacs,amsmath,amssymb]{revtex4-1}
\usepackage{lmodern}
\usepackage[T1]{fontenc}
\usepackage[latin9]{inputenc}
\usepackage{booktabs}
\usepackage{bm}
\usepackage{graphicx}
\usepackage{upgreek}
\usepackage{color}
\usepackage{multirow}

\makeatletter

\providecommand{\tabularnewline}{\\}

\newcommand{\MnSb}{Mn$_{1+\delta}$Sb}

\newcommand{\MnSbx}{Mn$_{1.13}$Sb}

\newcommand {\TC} {$T_{\mathrm{C}}$}

\newcommand {\TSR} {$T_{\mathrm{SR}}$}

\newcommand {\etal} {\emph{et al.}}

\makeatother

\usepackage{babel}
\begin{document}


\title{Influence of interstitial Mn on magnetism in room-temperature ferromagnet
\MnSb{}}

\author{A.~E.~Taylor}\email{{t}aylorae@ornl.gov}

\affiliation{Quantum Condensed Matter Division, Oak Ridge National Laboratory,
Oak Ridge, Tennessee 37831, USA}

\author{T.~Berlijn}

\affiliation{Center for Nanophase Materials Sciences and Computer Science and Mathematics Division, Oak Ridge National Laboratory, Oak Ridge, TN 37831-6494, USA}

\author{S.~E.~Hahn}

\affiliation{Neutron Data Analysis and Visualization Division, Oak Ridge National Laboratory,
Oak Ridge, Tennessee 37831, USA}

\author{A.~F.~May}

\affiliation{Materials Science and Technology Division, Oak Ridge National Laboratory,
Oak Ridge, Tennessee 37831, USA}

\author{T.~J.~Williams }
\affiliation{Quantum Condensed Matter Division, Oak Ridge National Laboratory,
Oak Ridge, Tennessee 37831, USA}

\author{L.~Poudel}
\affiliation{Quantum Condensed Matter Division, Oak Ridge National Laboratory,
Oak Ridge, Tennessee 37831, USA}

\affiliation{Department of Physics and Astronomy, University of Tennessee, Knoxville,
TN 37996, USA}

\author{S.~Calder}

\affiliation{Quantum Condensed Matter Division, Oak Ridge National Laboratory,
Oak Ridge, Tennessee 37831, USA}

\author{R.~S.~Fishman }
\affiliation{Materials Science and Technology Division, Oak Ridge National Laboratory,
Oak Ridge, Tennessee 37831, USA}

\author{M.~B.~Stone }

\author{A.~A.~Aczel }

\author{H.~B.~Cao}

\author{M.~D.~Lumsden}

\affiliation{Quantum Condensed Matter Division, Oak Ridge National Laboratory,
Oak Ridge, Tennessee 37831, USA}

\author{A.~D.~Christianson}

\affiliation{Quantum Condensed Matter Division, Oak Ridge National Laboratory,
Oak Ridge, Tennessee 37831, USA}

\affiliation{Department of Physics and Astronomy, University of Tennessee, Knoxville,
TN 37996, USA}

\pacs{75.50.Cc, 75.25.-j, 71.15.Mb, 78.70.Nx}

\begin{abstract}
We report elastic and inelastic neutron scattering measurements of the high-\TC{} ferromagnet \MnSb{}.
Measurements were performed on a large, $T_\mathrm{C}=434\,$K, single
crystal with interstitial Mn content of $\delta\approx0.13$. The
neutron diffraction results reveal that the interstitial Mn has a
magnetic moment, and that it is aligned antiparallel to the main Mn moment.
We perform density functional theory calculations including the interstitial
Mn, and find the interstitial to be magnetic in agreement with the diffraction data. The
inelastic neutron scattering measurements reveal two features in the
magnetic dynamics: i) a spin-wave-like dispersion emanating from ferromagnetic
Bragg positions (H$\,$K$\,$2$n$), and ii) a broad, non-dispersive
signal centered at forbidden Bragg positions (H$\,$K$\,$2$n$+1). The inelastic spectrum cannot
be modeled by simple linear spin-wave theory calculations, and appears
to be significantly altered by the presence of the interstitial
Mn ions. The results show that the influence of the interstitial
Mn on the magnetic state in this system is more important than previously understood. 
\end{abstract}
\maketitle

\section*{Introduction}

\MnSb{} is a high Curie temperature (\TC{}), highly anisotropic,
metallic ferromagnet. The observation of a ferromagnetic state in
\MnSb{} is unusual in its class of materials. Most Mn-alloys are
antiferromagnetic, and other 3$d$ transition metal mono-antimonides
are not ferromagnetic --- CrSb, FeSb and CoSb are antiferromagnetic,
TiSb is paramagnetic and NiSb is diamagnetic~\cite{motizuki_electronic_2009}. The ferromagnetic state
in \MnSb{} is highly sensitive to substitutions; for example, Cr-doping
quickly tunes the system towards an antiferromagnetic state~\cite{takei_magnetic_1963,kohnke_anisotropic_1996}.
In-fact substitutions of the cation, the anion, or presence of an
interstitial can each alter \TC{}, the magnetic anisotropy, or the
type of magnetic order~\cite{kohnke_anisotropic_1995,kohnke_anisotropic_1996,takei_magnetic_1963,chen_properties_1978,reimers_magnetic_1982,nakamura_121sb_1992,okita_crystal_1968,teramoto_existence_1968}.
This potential for tuning the properties in \MnSb{}, and closely
related Mn$_{1+\delta}$Bi, has attracted considerable attention
because the materials show promise as alternatives to rare-earth containing
permanent magnets, and as magneto-optic mediums~\cite{anand_effects_2014,carey_magnetic_1990,coehoorn_electronic_1985-1,continenza_coordination_2001,di_magneto-optical_1992,hong_magnetic_2013,kharel_spin_2011,zarkevich_anomalous_2014,ashizawa_temperature_2002}.

\begin{figure}
\includegraphics[clip,width=0.95\columnwidth]{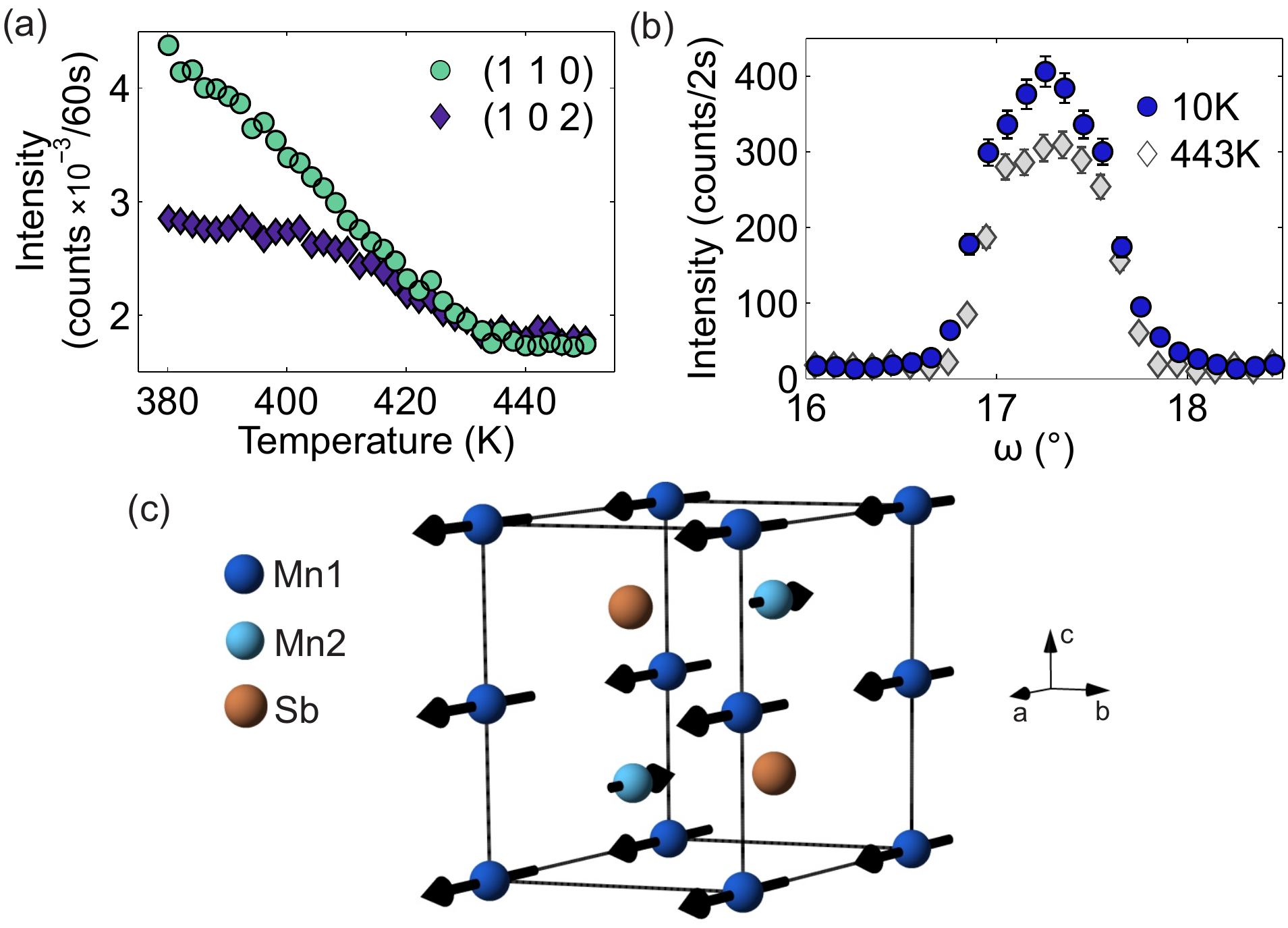}

\protect\caption{\label{fig:Diffraction}(Color online) (a) Temperature dependence of the peak intensity at the
$(1\,1\,0)$ (circles) and $(1\,0\,2)$ (diamonds) Bragg positions across \TC{}
measured on HB-3A. Error bars are smaller than the size of the markers.
(b) The $(1\,0\,3)$ Bragg peak measured at 10\,K {[}circles{]} and
443\,K {[}diamonds{]} as a function of goniometer angle $\omega$, which rotates the crystal in the scattering plane.
(c) One unit cell showing the low temperature ($T<T_\mathrm{SR}$) magnetic structure of \MnSb{}. The
interstitial sites are shown here as fully occupied. }
\end{figure}

The spontaneous magnetization in \MnSb{} is along the crystallographic $c$-axis at high temperatures. However, the anisotropy decreases on cooling and passes through zero at the spin reorientation temperature, \TSR{}, so that the magnetization is in-plane at low temperatures.
In nominally $\delta=0$ polycrystalline MnSb the \TC{} is as high as $T_\mathrm{C}=587\, K$~\cite{Guillaud_1949}. However, single crystal studies have been unable to produce the material without the inclusion of interstitial Mn ions~\cite{okita_crystal_1968}. These interstitial Mn ions, Mn2, are present in addition to the fully-occupied main site Mn ions, Mn1.
The Mn2 ions are hosted within the hexagonal NiAs crystal
structure as shown in Fig.~\ref{fig:Diffraction}(c). Compositions
in the range of $\delta\approx0.05-0.2$ have been found to be stable.

The presence of interstitials
significantly alters the properties of \MnSb{}. As $\delta$ increases,
the $a$ lattice parameter and unit cell volume increase, but the
$c$ lattice parameter, \TC{}, \TSR{} and the total magnetization
decrease~\cite{okita_crystal_1968,chen_properties_1978,teramoto_existence_1968}.
Studies combining magnetization measurements with chemical analysis~\cite{okita_crystal_1968,yamaguchi_magnetic_1976}
found that the interstitial Mn results in a reduction of the ferromagnetic ordering temperature as captured in a simple relationship between \TC{} and $\delta$:  
\begin{equation}
T_{\mathrm{{C}}}=(577-900\times\delta)\,{\mathrm{K}}.\label{eq:Tceq}
\end{equation}

The electronic structure and magnetic state of \MnSb{} have been investigated
by a number of authors via electronic structure calculations~\cite{coehoorn_electronic_1985-1,continenza_coordination_2001,katoh_electronic_1986,podloucky_electronic_1984,ravindran_magnetic_1999,vast_first-principles_1992}.
They found that MnSb is described as a metallic system
with localized magnetic moments, and show that the ferromagnetic state
is stabilized due to the significant hybridization between the Mn
3$d$ and Sb 5$p$ orbitals. All these calculations,
however, have failed to describe key experimental observations such
as the behavior of the magnetic anisotropy~\cite{ravindran_magnetic_1999},
and the effect of the interstitial Mn on the magnetic properties~\cite{coehoorn_electronic_1985-1}. Coehoorn \etal{}~\cite{coehoorn_electronic_1985-1}
simply discuss Mn2 as an electron donor, and were unable to
explain its influence on the magnetic properties. 

Given the importance of the interstitials in tuning the properties
of \MnSb{}, it is surprising that detailed theoretical investigations
of the effect of Mn2 are lacking. Experimental results are also limited,
and have focused on the empirical determination of the $\delta$ dependencies
of cell parameters, \TC{} and \TSR{}. This is probably due to a
prevailing view in the literature that there is no magnetic moment
associated with Mn2. This perspective originated from the analysis of polarized neutron
diffraction data~\cite{yamaguchi_polarized_1978,reimers_polarised_1983},
using a model with a complex aspherical magnetic form factor for Mn1.
A large asymmetry in the form factor allows the Mn1 site to contribute
magnetic intensity to Bragg reflections that would otherwise be forbidden for Mn1 scattering. In these analyses, low-$Q$
Bragg peaks had to be excluded to model the data. Analysis of powder neutron diffraction
data~\cite{bouwma_neutron_1971,yamaguchi_magnetic_1983}, x-ray
Compton profile measurements~\cite{nakamura_magnetic_1995}, and electronic
structure calculations~\cite{coehoorn_electronic_1985-1} raised questions as to how robust the aspherical magnetic form factor model is. 
Additionally, a density functional theory (DFT) calculation on the influence of
Pt doping in MnBi did include Mn at the interstitial site, created
via a vacancy on a main Mn site, and predicted a magnetic moment of
order $-3\,\upmu_{\mathrm{B}}$ associated with the interstitial site~\cite{kharel_spin_2011}. The absence
of Mn2 from most calculations may be attributable to the inherent
difficulty in including a partially occupied, disordered arrangement
of interstitial ions into calculations. The role of the interstitial
Mn in determining the magnetic properties of \MnSb{}, therefore,
is an open question which could provide key insights into the material
properties and tunability.

Here we show that the presence of the interstitial Mn ion has a strong
influence on the magnetic state of \MnSb{}, beyond the role of a
simple electron donor. We use neutron diffraction to measure a large
number of Bragg peaks from a $\delta=0.13$ single crystal sample at temperatures
above and below both \TC{} and \TSR{}. We find that the data is
described across the entire $Q$-range by a simple model including
a magnetic moment on the Mn2 site, aligned antiparallel to the moment on the Mn1 site.
We perform DFT calculations including an interstitial Mn, and find these to carry a sizable magnetic moment in agreement with our diffraction results. The DFT results are also consistent with 
the observed reduction in \TC{} due to the presence
of Mn2. We also investigate the magnetic dynamics of \MnSb{}
via inelastic neutron scattering (INS) measurements performed on a large single
crystal. We identify a spin-wave-like signal associated with the ferromagnetic Mn1 ions, and are able to partially reproduce its dispersion using a 
localized-moment Heisenberg model. In addition to the
spin-wave-like scattering, we also identify a broad, intense, magnetic response in
the inelastic spectrum. This signal is not observed in $\delta=0$ MnBi, and therefore is likely attributable to the presence of the interstitials in \MnSbx{}. These results highlight the strong influence that the presence of the interstitial Mn has on the magnetic properties of \MnSb{}.

\section*{Experimental Techniques}

The single crystal of \MnSb{} used in this investigation is the same
$\sim6\,$cm$^{3}$ crystal that was studied in Ref.~\cite{radhakrishna_inelastic-neutron-scattering_1996}.
Pieces taken from the large crystal were used for diffraction and
magnetization measurements. 

\begin{figure}
\includegraphics[clip,width=1\columnwidth]{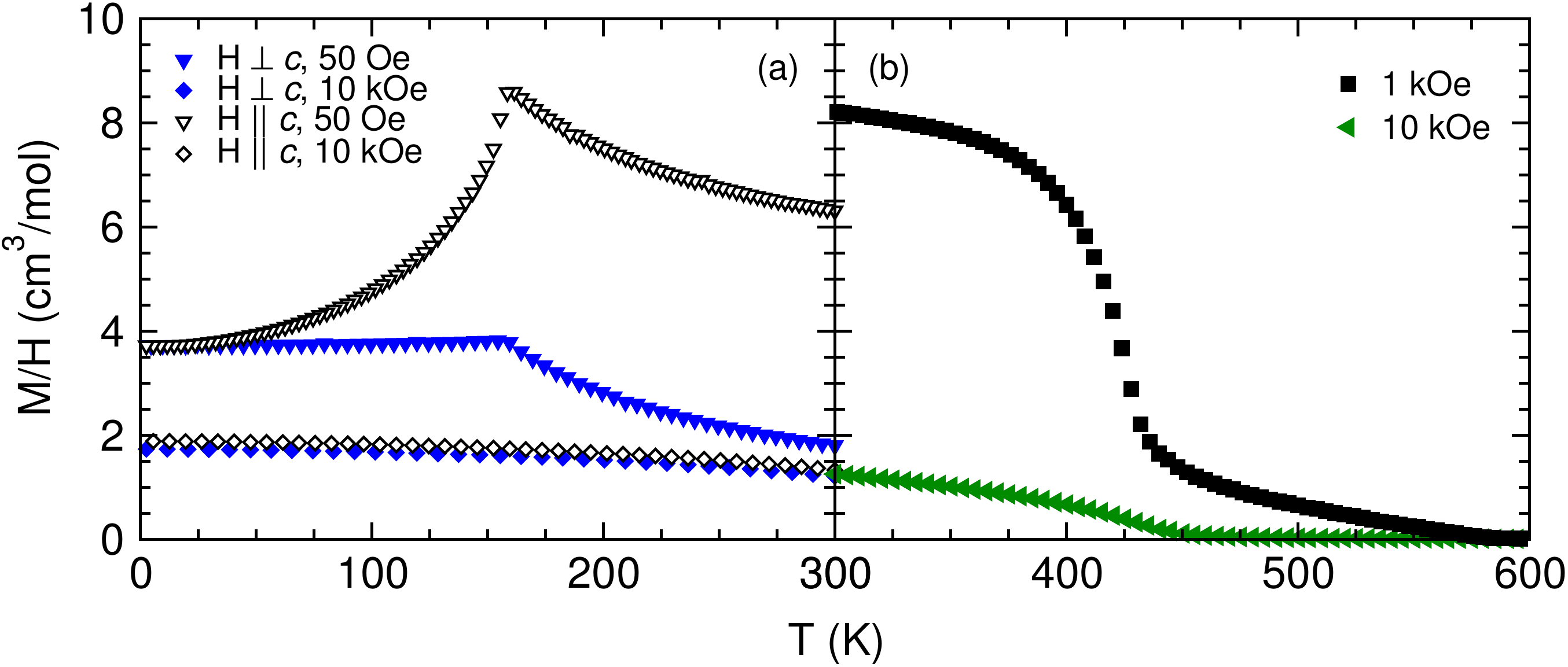}

\protect\caption{\label{fig:Magnetisation}(Color online) Magnetization data showing the two magnetic transitions, with (a) revealing the spin reorientation at 160\,K by examining data collected on an oriented crystal, and (b) demonstrating the bulk Curie temperature near 434\,K.  All of the data were collected while cooling in an applied field, except for the 10\,kOe data that were collected on the first warming measurement above 300K. }
\end{figure}

To determine the spin reorientation temperature, a piece of the crystal was aligned at room temperature to within ~15$^{\circ}$ of the easy axis ($c$-axis) using a permanent magnet.  The magnetization (\textbf{M}) 
was then measured in a Quantum Design Magnetic Property Measurement System, with applied fields \textbf{H} either parallel or perpendicular to this axis.  This measurement clearly reveals a spin reorientation temperature of $\sim160\,$K, as shown in Fig.~\ref{fig:Magnetisation}(a).  The same crystal was then used to obtain the Curie temperature of 434\,K,  see Fig.~\ref{fig:Magnetisation}(b). This high-temperature measurement employed the Sample Space Oven from Quantum Design, which utilizes thin quartz holders that precluded alignment of the crystal. Here, we have defined \TC{} as the intercept of the greatest tangent to the M/H versus $T$ data, with the greatest tangent being observed at 423\,K.  According to Eq.~\ref{eq:Tceq},
$T_\mathrm{C} = 434\,$K gives an interstitial content of 16$\,$\%, consistent with the
value predicted from the lattice parameters in Ref.~\cite{radhakrishna_inelastic-neutron-scattering_1996}.
With an applied field of 6\,T, the moment is essentially saturated at 2\,K, and for \textbf{H} $\perp$ $c$ we measure a saturated moment of 3.20$\,\upmu_B$/f.u., where f.u. is a formula unit \MnSbx{}. 
The high-temperature data shows a minor onset of magnetization magnetization at $T\sim580$\,K, see Fig.~\ref{fig:Magnetisation}(b). This is likely associated with a small close-to-stoichiometric MnSb impurity, or a small amount of an unknown phase.

Neutron diffraction measurements were performed on the HB-3A four-circle
diffractometer at the High Flux Isotope Reactor (HFIR) at Oak Ridge
National Laboratory (ORNL), using a neutron wavelength of 1.003$\,\mathrm{{\AA}}$.
Full datasets of more than 190 reflections were collected at temperatures
of 10\,K ($<$\TSR{}), 200\,K (\TSR{}$<200\mathrm{\,{K}}<$\TC{})
and 450\,K ($>$\TC{}), in a closed-cycle refrigerator (CCR). 450\,K data were counted for 2\,s per point
and 200\,K and 10\,K data were counted 5\,s per point. In addition, a set
of 10 peaks, which are predicted  by model 1  (see Results section)  to be highly sensitive to the direction of the interstitial moment with respect to the Mn1 site moment, were counted for up to 60\,s per point at all temperatures. 

INS measurements were performed on
the SEQUOIA spectrometer at the Spallation Neutron Source (SNS), ORNL
with the crystal aligned with (H$\,$0$\,$L) in the scattering plane.
Measurements were performed with an incident energy ($E_{\mathrm{i}}$)
of 150$\,$meV with chopper frequency of $f=300\,$Hz, and with $E_{i}=60\,$meV
and $f=180\,$Hz. This gave energy resolutions at the elastic line of 11\,meV and 4\,meV,
respectively. A CCR was
used to reach sample temperatures between 10$\,$K and 443$\,$K.
Full data sets were collected at 10$\,$K, 350$\,$K and 443$\,$K with
$E_{\mathrm{i}}=150\,$meV over at least a 130$^{\circ}$ range of sample rotation angles with
a step size of 0.5$^{\circ}$. Limited angular range data sets with
1$^{\circ}$ step size were collected at 40$\,$K, 100$\,$K, 125$\,$K,
150$\,$K, 170$\,$K, 200$\,$K, 250$\,$K and 300$\,$K. Measurements
were performed with $E_{\mathrm{i}}=60\,$meV at 10$\,$K and 350$\,$K
over an 80$^{\circ}$ range with a 2$^{\circ}$ step size. The angular
step data were combined, a slight misorientation of the crystal was
corrected for, and cuts through the data were performed using the Horace
software package~\cite{horace}. Additional INS measurements were
performed on the US/Japan Cold Neutron Triple-Axis Spectrometer (CTAX)
at HFIR, ORNL, using guide-open-80$^\prime$-open collimation and fixed final neutron
energy of 5$\,$meV.

\section*{Results}

\subsection*{Neutron diffraction}

Results from the neutron diffraction measurements are shown in Fig.~\ref{fig:Diffraction}. 
The temperature dependencies of $(1\,1\,0)$
and $(1\,0\,2)$ Bragg reflections across \TC{} are shown in Fig.~\ref{fig:Diffraction}(a). These reflections have structure factor contributions
from all three atoms in the unit cell. The intensities of these Bragg
peaks respond to \TC{}, indicating that they are
magnetic Bragg reflections, and confirming the identification of \TC{} from magnetization. Figure~\ref{fig:Diffraction}(b) shows
the $(1\,0\,3)$  Bragg reflection at 443\,K and 10\,K. The Mn1 site 
in the NiAs crystal structure has the reflection condition $l=2n$,
where $n$ is an integer. Therefore, the $(1\,0\,3)$ peak has zero nuclear
contribution to its structure factor from Mn1, and also zero
magnetic contribution if the magnetic form factor of Mn1 is
spherical. Both Mn2 and Sb ions contribute scattering intensity
to the $(1\,0\,3)$ reflection.

To determine the origin of the increased intensity at low temperature
of $l=\mathrm{odd}$, $(h-k)\neq 3n$ Bragg peaks, we investigated
two possible models for the magnetic state in \MnSb{} for all reflections
collected. Both models use the NiAs structure, hexagonal space group
$P6_{3}/mmc$ (No. 194), which has two formula units of \MnSb{} per unit cell, with Mn1 on the 2$a$ Wyckoff site (0\,0\,0), 
Sb on the 2$c$ Wyckoff site $(\frac{1}{3}\,\frac{2}{3}\,\frac{1}{4})$, and Mn2 on the 2$d$
Wyckoff site $(\frac{1}{3}\,\frac{2}{3}\,\frac{3}{4})$. We use standard Miller indices (h k l) to index the reflections. Note that because 
of the hexagonal symmetry the Miller-Bravais notation can be used to illustrate equivalent reflections, given by (h k i l) where i=-(h+k) 
and permutations of h k i give equivalent reflections~\cite{frank_millerbravais_1965}.
Recent x-ray diffraction results from the closely related
compound MnBi identify a slight distortion from hexagonal symmetry
below \TSR{}, however the resolution of our neutron diffraction measurement is
not sufficient to detect a distortion of this size~\cite{mcguire_symmetry-lowering_2014}.
The difference between the two models we investigate concerns the
magnetic component of scattering below \TC{}.

The first model comprises a ferromagnetic arrangement of Mn1
moments, with Mn2 moments aligned antiparallel to Mn1. 
A depiction of this model is shown in Fig.~\ref{fig:Diffraction}(c)
for $T<$\TSR{}, i.e. with spins aligned within the $a$-$b$ plane.
For $T>$\TSR{} the Mn1 and Mn2 moments are aligned along the $c$-axis, but still antiparallel to each other. 
The 3$d$-5$p$ hybridization is expected to induce a small
moment on the Sb site antiparallel to Mn1~\cite{coehoorn_electronic_1985-1}. However,
considering the expected moment size and the steep magnetic form factor for 
Sb 5$p$ electrons~\cite{yamaguchi_polarized_1978}, the overall contribution of Sb to
the magnetic diffraction pattern is expected to be small. Therefore we do not include magnetic
 intensity from Sb in the fitting for either model 1 or model 2, consistent with previous reports~\cite{bouwma_neutron_1971,yamaguchi_polarized_1978,reimers_polarised_1983}.
As Sb and Mn2 have the same reflection conditions, this may
have the effect of slightly increasing the moment size assigned to
the interstitial Mn. 

\begin{figure}
\includegraphics[clip,width=0.95\columnwidth]{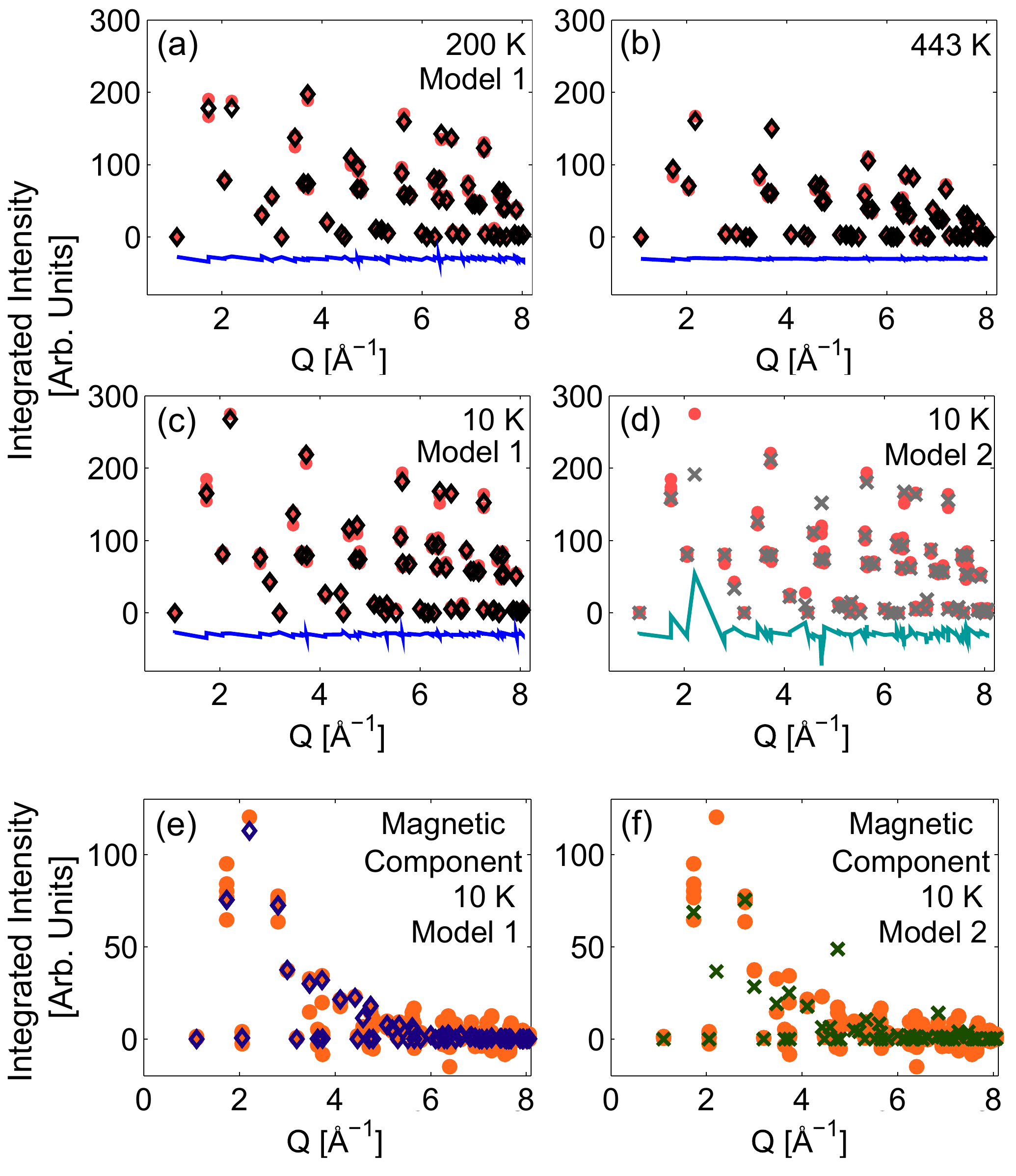}

\protect\caption{\label{fig:Fullprof}(Color online) Results of model 1 fitted in FullProf compared
to (a) 200\,K, (b) 443\,K and (c) 10\,K  data. (d) Model 2 compared to 10\,K data. 
The data points (red circles) are the integrated
intensities of peaks in scans similar to those shown in Fig.~\ref{fig:Diffraction}(b) 
for all measured Bragg positions. The results predicted from model
1 (open diamonds), and model 2 (crosses) are shown. The difference
between the data and models (solid lines) have been offset from zero for clarity.
(e) and (f) Show the extracted magnetic component of the scattering. The data minus the calculated structural part of the scattering are plotted (orange circles), along with the calculated magnetic component  of the scattering for model 1 (open diamonds) and model 2 (crosses). }
\end{figure}

The model was fit to the data using Rietveld refinement in
the FullProf software suite~\cite{rodriguez-carvajal_recent_1993}. 
For a random distribution of interstitial Mn ions, the average structure 
can be modeled by assuming a uniform distribution of Mn2 on every 2$d$
Wyckoff site and scaling the scattering intensity by the occupancy
$\delta$. The atomic displacement parameters for Sb and
Mn2 were constrained to be equal, as they contribute to the
same reflections. The initial fit was performed against the $T=450$\,K (i.e. $T>T_\mathrm{C}$)
data set, allowing us to determine the interstitial content of the
sample without the influence of magnetic Bragg scattering. The standard empirical correction for extinction used in FullProf was utilised in this refinement~\cite{rodriguez-carvajal_recent_1993}, resulting in an extinction coefficient of 580(60) which gives a maximum extinction correction of 68\,\% on the strongest Bragg peak. Using this fitting procedure the value of the interstitial content was determined to be $\delta=0.13(1)$, which is reasonably close to the value estimated from
comparison of the magnetization measurement (see Fig.~\ref{fig:Magnetisation})
to Eq.~\ref{eq:Tceq}. For subsequent fits of the magnetic model
against 10\,K and 200\,K data sets the interstitial content and
the extinction parameter were kept fixed. The magnetic form factor of Mn$^{2+}$ was used~\cite{reimers_polarised_1983,dianoux_neutron_2003}. For the 10\,K data set,
the model included three domains in equal proportions with spins along
$a$, $b$, and {[}-1 -1 0{]} directions respectively. The results
of these fits are summarized in Fig.~\ref{fig:Fullprof}(a), (b) and~(c) and
Table~\ref{tab:Diffraction}. Excellent agreement between data and
model 1 is found.

\begin{table*}
\begin{tabular}{cccccccccc}
\toprule 
Temperature (K) & $a$ ($\mathrm{{\AA}}$) & $c$ ($\mathrm{{\AA}}$) & Vol. ($\mathrm{{\AA}}^{3}$) & $M{}_{\mathrm{Mn1}}$ ($\upmu_{\mathrm{B}}$) & $M{}_{\mathrm{Mn2}}$ ($\upmu_{\mathrm{B}}$) & ext & scale & $\chi^{2}$ & $R_{\mathrm{F}}$\tabularnewline
\midrule 
450\,K & 4.193(3) & 5.773(4) & 87.9(1) & -- & -- & 580(60)  & 346(7) & 0.325 & 3.79 \tabularnewline
200\,K & 4.195(6) & 5.723(8) & 87.2(2) & 2.92(4) $[\parallel c]$$\hspace{1em}$ & -1.4(3) $[\parallel c]$$\hspace{1em}$ & 580 &  368(5) & 2.45 & 4.05\tabularnewline
10\,K & 4.189(5) & 5.690(7) & 86.5(2) & 3.54(2) $[\perp c]$$\hspace{1em}$ & -2.5(1) $[\perp c]$$\hspace{1em}$ & 580 &  372(6) & 2.46 & 4.55\tabularnewline
\bottomrule
\end{tabular}

\protect\caption{Results from Rietveld refinements fitting model 1 to the data.
Vol. is volume of the unit cell, $M_{\mathrm{Mn1}}$ and $M_{\mathrm{Mn2}}$
are the magnitudes of the magnetic moments on Mn1 and Mn2 sites respectively, ext and scale are the extinction and scale factors and 
$\chi^{2}$ and $R_{\mathrm{F}}$ are the statistical agreement factors
determined by FullProf. $\chi^{2}$ is low for the 450\,K data set
because of the shorter collection time used for this data set. The
errors on the moment sizes are the estimated standard deviation calculated
by FullProf. \label{tab:Diffraction}}
\end{table*}

To highlight the magnetic component of the scattering, Fig.~\ref{fig:Fullprof}(e) shows a comparison of the 10K data minus the calculated structural component of the scattering, with the calculated magnetic contribution to the scattering. The intensity of the magnetic contribution is seen to decrease with $Q$ for both data and calculation, as expected due to the magnetic form factor. The effects of the structure factor and polarization factor, which selects scattering from only moments perpendicular to $\bm{Q}$,  in the neutron scattering cross section are present in this data, which is why there is not an exact form factor dependence. In addition, some forbidden Bragg positions were measured and appear at zero integrated intensity. Attempts to fit an interstitial Mn moment aligned parallel to Mn1, instead of antiparallel, were unsuccessful. The model with an antiparallel moment
on the Mn2 site describes the data very well.

The second model is based on that proposed by Haneda \etal{}~\cite{haneda_electronic_1977}
for MnAs and discussed in detail for MnSb by Yamaguchi \etal{}~\cite{yamaguchi_polarized_1978},
in which Mn2 does not have a magnetic moment associated with
it. In this case, they explain the observed magnetic scattering at $(1\,0\,3)$
and other $l=$odd Bragg positions~\cite{bouwma_neutron_1971,yamaguchi_polarized_1978,reimers_polarised_1983}
using a highly aspherical magnetic form factor for the main site Mn.
Having an aspherical magnetization density breaks the symmetry conditions
that normally result in systematic absences for $l=\mathrm{odd}$ reflections from the 2$a$ Wyckoff site, allowing intensity at $l=\mathrm{odd}$, $(h-k)\neq3n$ positions. The model includes
the combined scattering intensity for structural contributions from
all ions, plus the magnetic intensity from the Mn1 ions for
all reflections. We have added to model 2 the same extinction correction as used in model 1. The overall structure factor resulting from the Yamaguchi
model, with adjusted interstitial content and electron occupancies
to match our data, is compared to the 10\,K data in Fig.~\ref{fig:Fullprof}(d). For this model we find $\chi^2 = 2.49$ and $R_\mathrm{F} = 6.21$, compared with $\chi^2 = 2.46$ and $R_\mathrm{F} = 4.55$ for model 1 at 10\,K (Table~\ref{tab:Diffraction}). The purely magnetic component of model 2 is shown in  Fig.~\ref{fig:Fullprof}(f), along with the difference between the data and calculated structural component. 
In this case there is significant deviation between the model and the data over a large range of $Q$, with the intensities of only some magnetic peaks being replicated. Therefore model 2 is not as effective as model 1 in describing the results of our neutron diffraction experiment.
This will be examined further in the Discussion Section. 

\subsection*{DFT calculations}

\begin{figure}
\centering
\includegraphics[width=1\columnwidth]{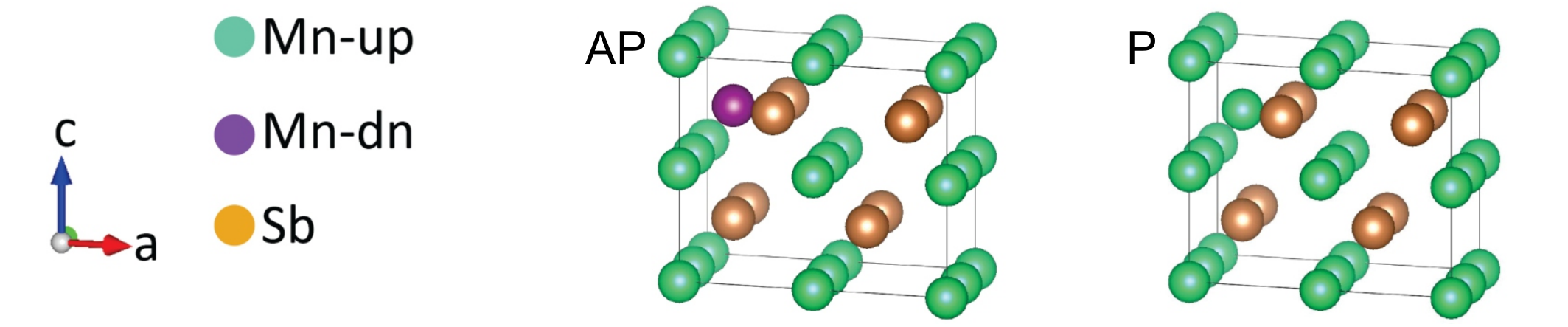}
\resizebox{\columnwidth}{!} {
\begin{tabular}{|c|c|cccccc|}
\hline
$U$  & & $a$($\mathrm{{\AA}}$) & $c$($\mathrm{{\AA}}$) & $M_\mathrm{Mn1}$($\upmu_{\mathrm{B}}$)  & $M_\mathrm{Mn2}$($\upmu_{\mathrm{B}}$)  & $M_\mathrm{Sb}$($\upmu_{\mathrm{B}}$) & E(meV)   \tabularnewline
\hline
\multirow{2}{*}{0} &  AP & 4.164 & 5.558 & 3.208 & -2.570 & -0.105 & 0 \tabularnewline 
 & P & 4.133 & 5.523 & 3.104 & 0.909 & -0.125 & 117 \tabularnewline
\hline
\multirow{2}{*}{2} & AP & 4.280 & 5.969 & 4.052 & -3.810 & -0.131 & 0 \tabularnewline 
 & P & 4.311& 5.942 & 4.082 &  3.915 & -0.123  & 281 \tabularnewline
\hline
\multirow{2}{*}{4} & AP & 4.338 & 6.231 & 4.386 & -4.261 & -0.166 & 0 \tabularnewline 
 & P & 4.352 & 6.254 & 4.417 & 4.284 & -0.155 & 18 \tabularnewline
\hline
\multirow{2}{*}{6} & AP & 4.388 & 6.353 & 4.566 & -4.518 & -0.194 & 0 \tabularnewline 
 & P & 4.393 & 6.365 & 4.575 & 4.483 & -0.199 & -119 \tabularnewline
\hline
\multirow{2}{*}{8} & AP & 4.425 & 6.443 & 4.693 & -4.683 & -0.210 & 0 \tabularnewline 
 & P & 4.430 & 6.441 & 4.693 & 4.633 & -0.223 & -144 \tabularnewline
\hline
\end{tabular}
}
\caption{(Color online) The optimized lattice constants, the average host Mn1 moment, the interstitial Mn2 moment, the average Sb moment, and the total energy per interstitial Mn2 atom of the parallel (P) and antiparallel (AP) configuration of Mn2 calculated within the PBE+U approximation for U=0,2,4,6 and 8\,eV.}
\label{tom1}
\end{figure}

To gain a better understanding of the influence of the Mn interstitials on \MnSb{} we performed DFT calculations including an interstitial in the unit cell (see~\cite{dft} for technical details). First we readdress the question of whether the interstitial Mn has a magnetic moment, and if so how it is directed with respect to the ferromagnetically aligned moments of the host Mn atoms. To this end we considered the 2$\times$2$\times$1 supercell Mn$_9$Sb$_8$ depicted in Fig.~\ref{tom1}, which contains a single interstitial Mn atom.
To account for the influence of Coulomb interactions among the Mn-3$d$ electrons we use the PBE+U approximation, and vary the Hubbard $U$ parameter between 0 and 8\,eV. 
For all cases, we find that it is not possible to stabilize a configuration in which the interstitial Mn is non-magnetic.
The tables in Fig.~\ref{tom1} show that for $U=0$, 2 and 4\,eV the configuration with Mn2 aligned parallel to Mn1 is higher in energy than the configuration in which it is aligned antiparallel, consistent with the results from neutron diffraction. For $U=6$ and 8\,eV, however, the configuration with the parallel alignment of the interstitial Mn is energetically favored.  It is reasonable to expect that the Coulomb $d$-$d$ interactions are at the lower end of the scale, given that the very large wave functions of the Sb 5$p$ electrons can effectively screen the transition metal interactions. Related to this, we find for all cases that the system remains metallic, even for a Hubbard $U$ of 8\,eV.

\begin{figure}[tbp]
\centering
\includegraphics[width=1\columnwidth]{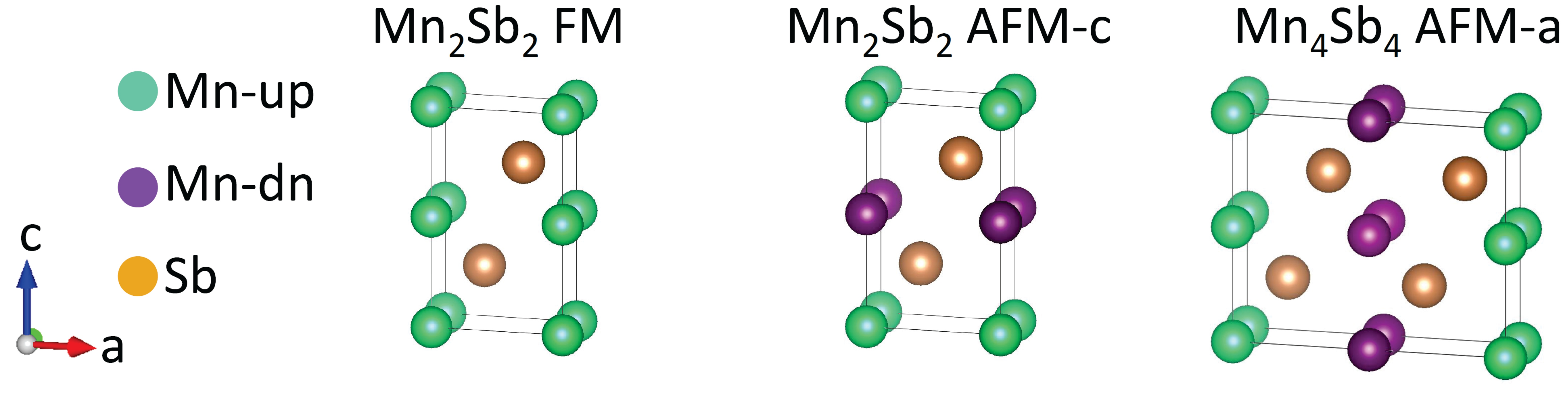}
\vspace{2mm}
\resizebox{0.98\columnwidth}{!} {
{\large (a)}\hspace{5mm}\begin{tabular}{|c|cccc|}
\hline
Without Mn2 & $a$($\mathrm{{\AA}}$) & $c$($\mathrm{{\AA}}$) & $M_\mathrm{Mn1}$($\upmu_{\mathrm{B}}$)  & E(meV)  per \tabularnewline
 & & & & Mn$_2$Sb$_2$ \tabularnewline
\hline
Mn$_2$Sb$_2$ FM & 4.189 & 5.998 & 4.027 & 0 \tabularnewline 
Mn$_2$Sb$_2$ AFM-c  & 4.189 & 5.998 & 3.974 & 286 \tabularnewline
Mn$_4$Sb$_4$ AFM-a & 4.189 & 5.998 & 3.957 & 326 \tabularnewline
\hline
\end{tabular}
}
\newline
\vspace{1mm}
\newline
\resizebox{0.98\columnwidth}{!} {
{\large (b)}\hspace{5mm}\begin{tabular}{|c|cccc|} 
\hline
Without Mn2 & $a$($\mathrm{{\AA}}$) & $c$($\mathrm{{\AA}}$) & $M_\mathrm{Mn1}$($\upmu_{\mathrm{B}}$)  & E(meV)  per \tabularnewline
with strain & & & & Mn$_2$Sb$_2$ \tabularnewline
\hline
Mn$_2$Sb$_2$ FM & 4.280 & 5.969 & 4.074 & 0 \tabularnewline 
Mn$_2$Sb$_2$ AFM-c  & 4.280 & 5.969 & 4.047 & 286 \tabularnewline
Mn$_4$Sb$_4$ AFM-a & 4.280 & 5.969 &  4.028 & 339 \tabularnewline
\hline
\end{tabular}
}
\includegraphics[width=1\columnwidth]{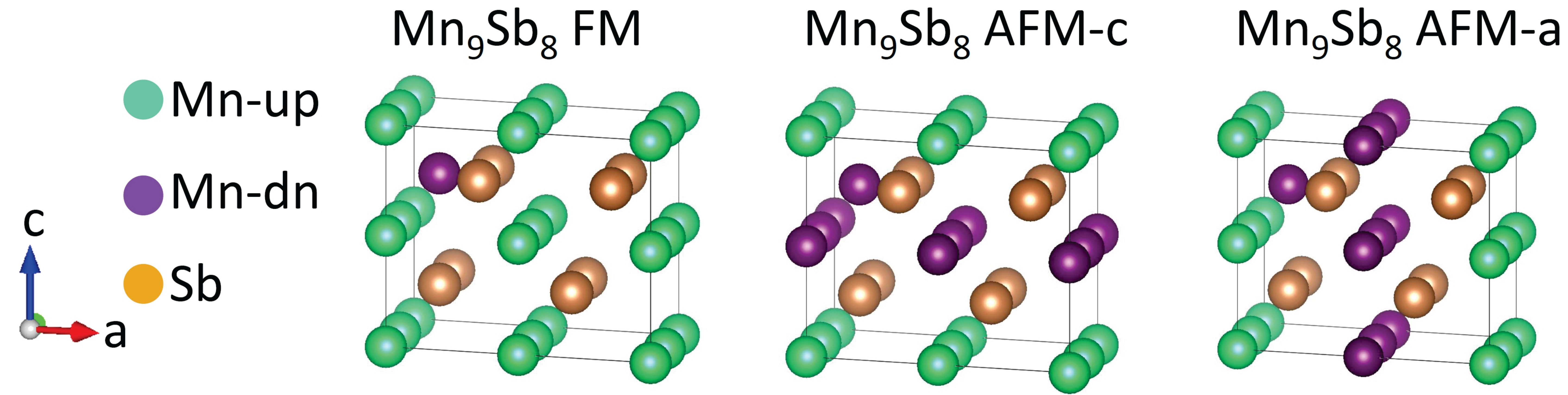}
\vspace{-0.7cm}
\begin{flushleft}
{\large (c)}
\end{flushleft}
\vspace{-0.2cm}
\centering
\resizebox{\columnwidth}{!} {
\begin{tabular}{|c|ccccc|}
\hline
With Mn2 & $a$($\mathrm{{\AA}}$) & $c$($\mathrm{{\AA}}$) & $M_\mathrm{Mn1}$($\upmu_{\mathrm{B}}$) &  $M_\mathrm{Mn2}$($\upmu_{\mathrm{B}}$)  & E(meV)  per \tabularnewline
 & & & & & Mn$_2$Sb$_2$ \tabularnewline
\hline
Mn$_9$Sb$_8$ FM & 4.280 & 5.969 & 4.052 & -3.810 & 0 \tabularnewline 
Mn$_9$Sb$_8$ AFM-c  & 4.280 & 5.969 & 4.043 & -3.780 &  212 \tabularnewline
Mn$_9$Sb$_8$ AFM-a & 4.280 & 5.969 & 4.018 & -3.736 & 306 \tabularnewline
\hline
\end{tabular}
}
\caption{(Color online) The lattice constants, the average of the absolute value of the host Mn1 moments, the interstitial Mn2 moment and the total energy per Mn$_2$Sb$_2$ unit cell of the FM, AFM-c and AFM-a configurations for the cases without ((a) and (b)) and with ((c)) an interstitial Mn2 atom, calculated within the PBE+U approximation with U=2\,eV. The lattice constants and internal parameters for the cases (a) and (c) are obtained from optimizing the Mn$_2$Sb$_2$-FM/Mn$_9$Sb$_8$-FM configuration, respectively. In (b) the lattice constants from the configuration with Mn2 are used in the unit cells without Mn2, giving effective strained lattice constants. }
\label{tom2}
\end{figure}

Next we examine the effect of the Mn interstitials on the ferromagnetic configuration of the host Mn moments.
To investigate the stability of the ferromagnetic ground state (FM) we compare its energy with that of two higher energy configurations in which the atomic positions remain unchanged, but the host Mn moments assume an antiferromagnetic configuration either along the $c$-axis (AFM-c) or the $a$-axis (AFM-a) as shown in Fig.~\ref{tom2}. 
Comparing Fig.~\ref{tom2}(a) and~(c), the presence of the Mn interstitial lowers the energy difference between the FM ground state and the high energy AFM configurations.
The energy difference between the FM and the AFM-c configuration reduces quite significantly from 286 to 212\,meV and the energy difference between the FM and the AFM-a configuration decreases from 326 to 306\,meV. 
The results in Fig.~\ref{tom2} are for a $U = 2\,$eV, but for a $U$ of 8\,eV we find that this qualitative conclusion remains unchanged~\cite{U8ev}.

It is important to note that the simulations in Fig.~\ref{tom2}(a) and~(c) include the indirect influence of Mn2 via their induced changes in the lattice constants. 
From comparing the lattice parameters between the undoped and doped system in Fig. \ref{tom2}(a) and~(c) we see that, just like in the experimental observations~\cite{teramoto_existence_1968}, the Mn interstitials tend to increase the $a$ lattice parameter while decreasing the $c$ lattice parameter.
This leads to the question of whether the change in lattice parameters is solely responsible for the damaging influence of the Mn interstitials on the FM state. 
To address that question, we present in Fig.~\ref{tom2}(b) the stability of the FM state again, for a system which has the strained lattice parameters of the doped system but does not include the Mn interstitials.
The energy differences of this undoped strained system shown in  Fig.~\ref{tom2}(b)  differ significantly from those in the doped system shown in Fig.~\ref{tom2}(c).
This illustrates that Mn2 affects the ferromagnetic state not just via the change of lattice parameters, suggesting that the magnetic exchange interactions between the host and the interstitial Mn moments and/or doping effects play an important role. Previous attempts to explain the magnetic properties of \MnSb{} just via doping effects were not successful~\cite{coehoorn_electronic_1985-1}. 

\subsection*{Inelastic neutron scattering}

\begin{figure*}
\includegraphics[clip,width=1.9\columnwidth]{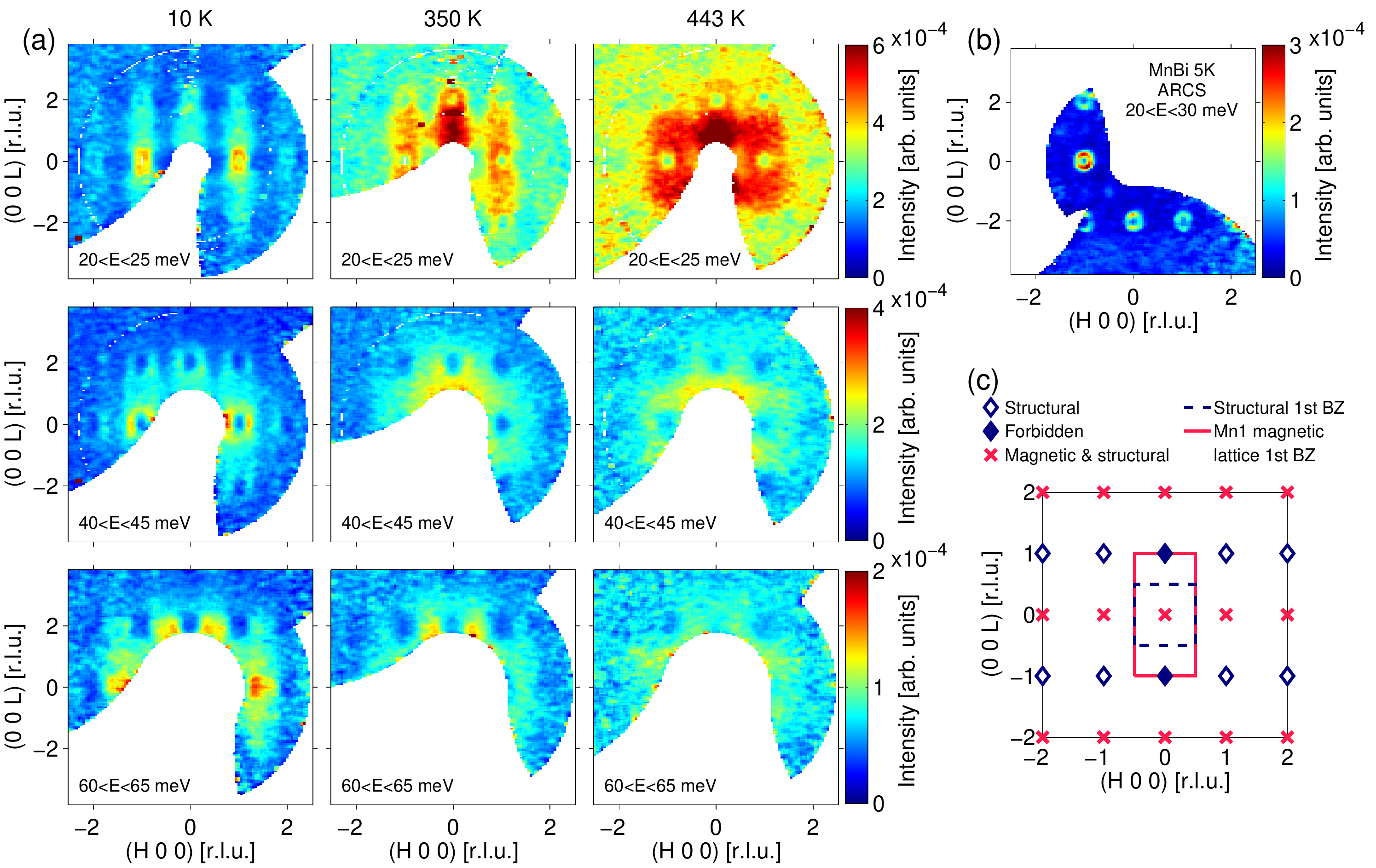}

\protect\caption{\label{fig:Const_E_slices}(Color online) (a) Neutron scattering intensity maps of \MnSb{}
in the HL-plane. The data were acquired on SEQUOIA at 10\,K, 200\,K
and 443\,K, as indicated, with $E_{\mathrm{i}}=150\,$meV. Slices
were averaged over the energy ranges indicated and over $\pm0.1$
reciprocal lattice units {[}r.l.u.{]} in the (0.5$\xi$ -$\xi$ 0)
direction (perpendicular to $(\mathrm{H}\,0\,0)$ and $(0\,0\,\mathrm{L})$). 
The data on interstitial-free MnBi in (b) was taken on the ARCS spectrometer at the SNS, at a temperature of 5\,K and with $E_{\mathrm{i}}=80$\,meV, courtesy of Ref.~\cite{williams_unpub}. (c) A map of reciprocal space for \MnSb{} indexed on the structural unit cell. The  first Brillouin Zones (BZs) of the structural unit cell (dashed line), and the magnetic unit cell of the Mn1 ions (solid line) are shown. Structural (diamonds and crosses) and Mn1 magnetic lattice (crosses) Bragg positions are indicated, with systematically absent reflections shown as closed diamonds.}
\end{figure*}

\begin{figure}
\includegraphics[clip,width=0.95\columnwidth]{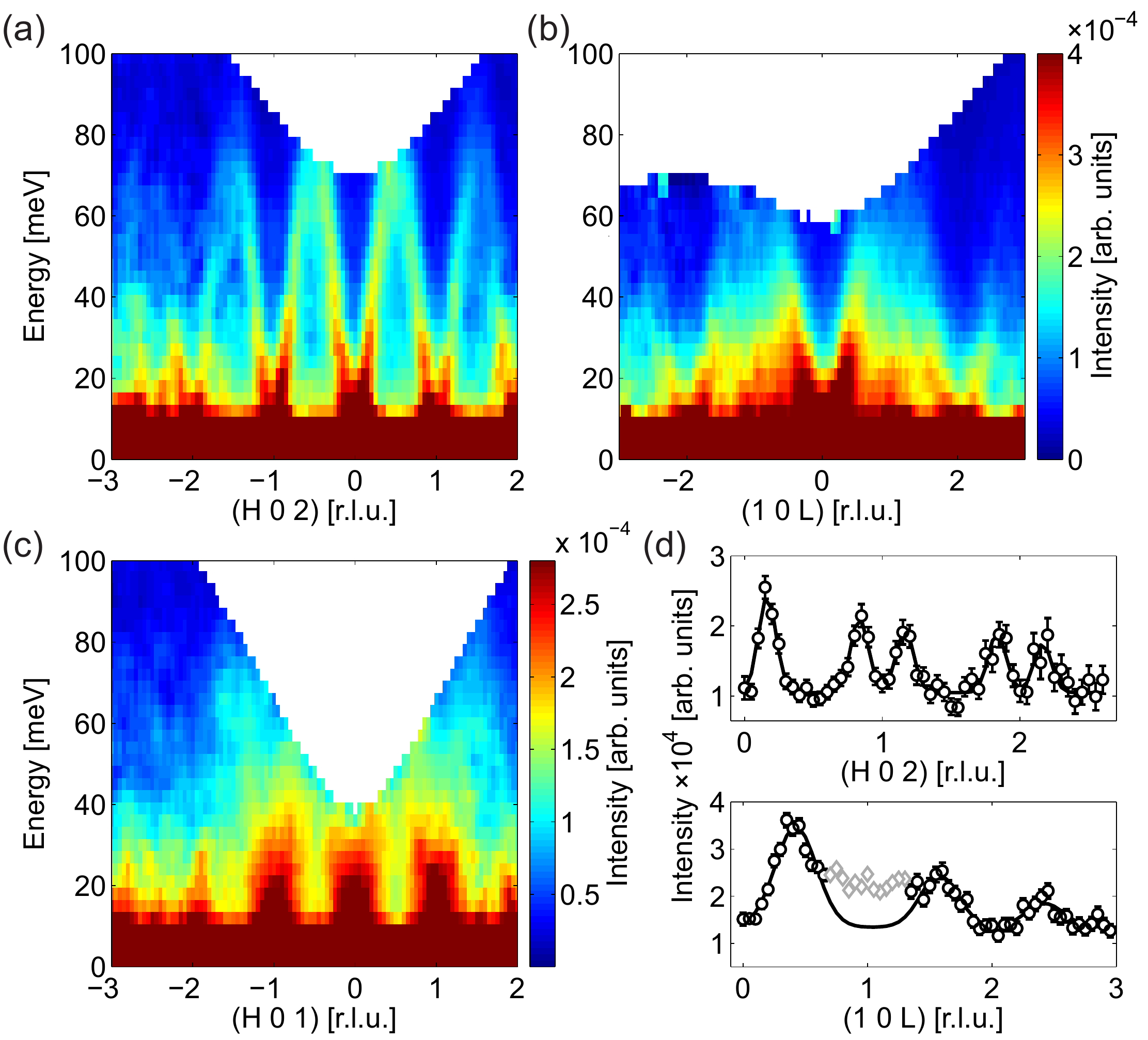}

\protect\caption{\label{fig:Dispersions}(Color online) (a-c) Neutron scattering intensity
maps for \MnSb{} from data taken on SEQUOIA with $E_{\mathrm{i}}=150\,$meV
at 10\,K. The data were averaged over $\pm0.1\,$r.l.u. in the two $\bm{Q}$-space directions perpendicular
to the $x$-axis in each case. (d) Markers show constant-energy cuts through the data
shown in (a) and (b), averaged over 20--25\,meV. Solid lines show
the result of fitting Gaussians plus a flat background to the data. The gray diamonds in
the $(1\,0\,\mathrm{L})$ cut show the region of data excluded from
the fit, as this region is dominated by the broad $(0\,0\,1)$ type
scattering from the $(1\,0\,1)$ position. }
\end{figure}

To investigate the magnetic state in \MnSbx{} further, we performed
INS experiments to probe the magnetic dynamics. An overview of the
results from SEQUOIA is given in Figs.~\ref{fig:Const_E_slices}(a) and~\ref{fig:Dispersions}, 
which show HL-plane and $\bm{{Q}}E$-plane
color maps, respectively, as well as constant-energy cuts through the
data in Fig.~\ref{fig:Dispersions}(d). 
These inelastic scattering spectra contain contributions from both magnetic and phonon scattering. However, the phonon scattering contribution is effectively negligible in the results we present. Mn and Sb have nuclear scattering cross sections of opposite sign, therefore their scattering contributions largely cancel-out and as a result the acoustic phonon scattering from the sample is weak. Additionally, the phonon intensity follows a $Q^2$ dependence and is therefore weak in the low-$Q$ zones we investigate.  In the low temperature data the phonon scattering is suppressed by the Bose-population factor in the energy range investigated, whereas at higher temperature phonons (attributable to both sample and background contributions) simply result in an overall increase in the background scattering, particularly at low energies, as evident in Fig.~\ref{fig:Const_E_slices}(a). Therefore the results presented are attributable to the magnetic dynamics of \MnSb{}.

\begin{figure}[tb]
\includegraphics[clip,width=0.6\columnwidth]{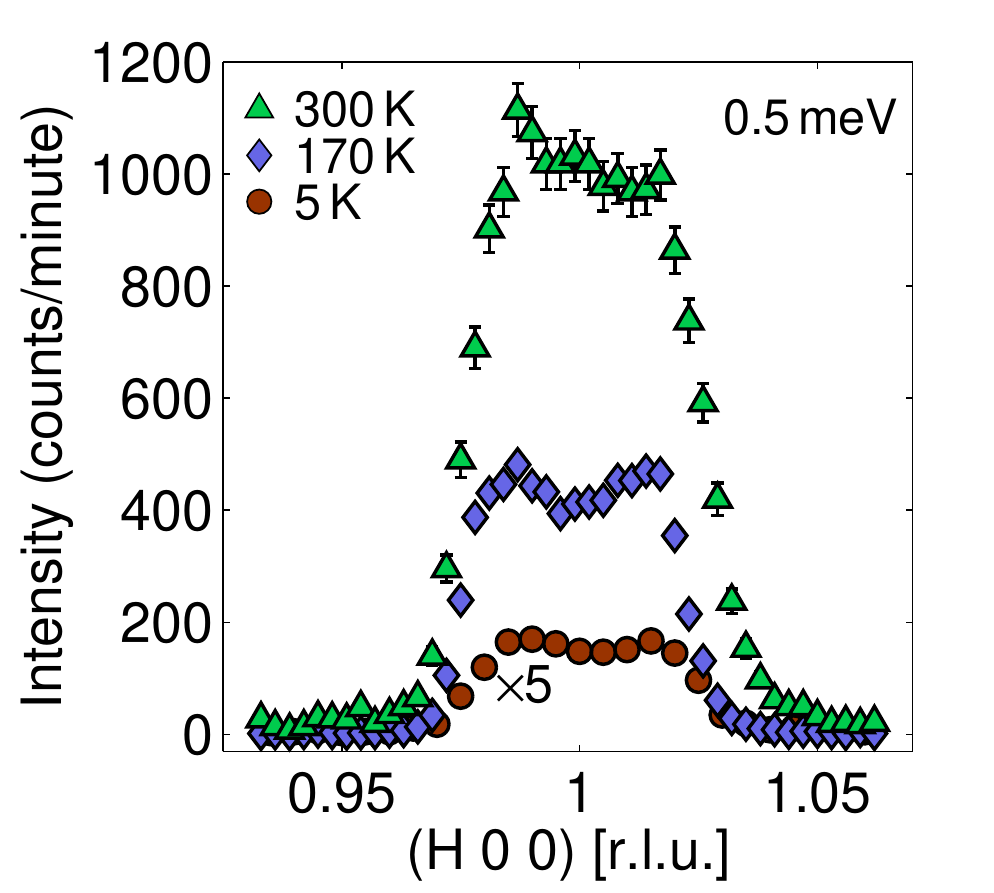}

\protect\caption{\label{fig:CTAX}(Color online) Low energy data measured on CTAX along the (H$\,$0$\,$0)
direction through ferromagnetic position (1$\,$0$\,$0). Measurements
were made with energy transfer of 0.5\,meV at 5, 170 and 300\,K, as indicated. The data have been
normalized to number of counts per minute. The 5\,K data has been multiplied by 5 for clarity.  }
\end{figure}

\begin{figure}
\includegraphics[clip,width=0.95\columnwidth]{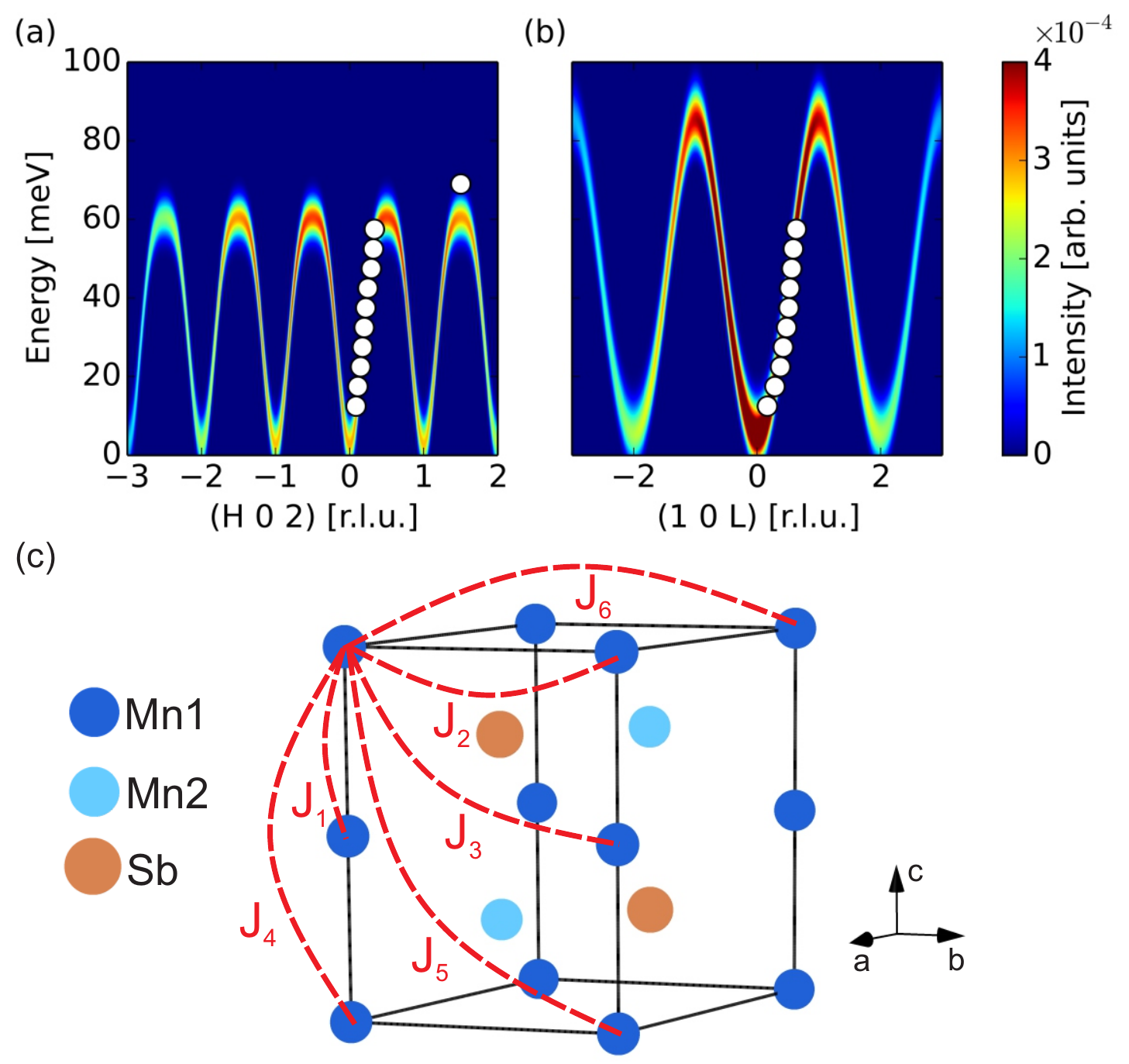}

\protect\caption{\label{fig:SpinWaveCalculations}(Color online) Spin-wave dispersion calculated for the H-direction (a) and L-direction (b), with calculated intensities corresponding to (H 0 2) and (1 0 L) directions, respectively. The white circles are the points extracted from cuts through the data (see Fig.~\ref{fig:Dispersions}(d) and text) that were used to fit the model parameters. (c) Depiction of exchange constants used in the Heisenberg model, from 1st to 6th nearest neighbors for Mn1 ions.}
\end{figure}

In Fig.~\ref{fig:Const_E_slices}(a) rings of scattering in the HL plane are seen dispersing with energy
out of the ferromagnetic positions such as $(1\,0\,0)$ and $(0\,0\,2)$; compare to Fig.~\ref{fig:Const_E_slices}(c) which depicts the $(\mathrm{H}\,0\,\mathrm{L})$ reciprocal lattice plane. Only L = even positions are magnetic zone centers because the Mn1 magnetic lattice is half the size of the structural lattice along the $c$-axis, see Fig.~\ref{fig:Diffraction}(c). 
The rings persist above both \TSR{} and \TC{}. The intensity of the rings is rapidly suppressed with increasing
$Q$ at all temperatures, and it persists up to energies of $\sim70$\,meV
(see Fig.~\ref{fig:Dispersions}(a)), well above the phonon cutoff
of the sample $\sim30$\,meV, indicating the magnetic origin of the
signal. We observe this signal down to low energies in the data from our measurement on CTAX, Fig.~\ref{fig:CTAX}.
This shows that the dispersion of the magnetic signal along H can already be observed out of $(1\,0\,0)$  with an energy transfer of 0.5\,meV, see Fig.~\ref{fig:CTAX}. This dispersive signal is reasonably reproduced by a spin-wave Heisenberg model for Mn1 ions, presented in Fig~\ref{fig:SpinWaveCalculations}, discussed further below, and a qualitatively similar signal is observed in MnBi~\cite{williams_unpub}, which contains no interstitial Mn, see Fig.~\ref{fig:Const_E_slices}(b). 


In the inelastic spectra, in addition to the spin-wave type signal, a broad signal is evident
between the rings in the HL plane at all temperatures, see Fig.~\ref{fig:Const_E_slices}(a).
This broad signal was not observed in MnBi, corroborating that it is
a distinct feature of the excitation spectrum of \MnSb{}, Fig.~\ref{fig:Const_E_slices}(b). The broad
signal is centered on $(0\,0\,1)$ and equivalent positions. $(0\,0\,1)$
is not an allowed Bragg reflection in this system (see Fig.~\ref{fig:Const_E_slices}(c)), and no Bragg peak
was observed at this position in any of the measurements we performed. 
Figure~\ref{fig:Dispersions}(c) highlights that this signal is broad but
localized in $Q$-space, and seemingly non-dispersive. Again, the intensity
suppression with $Q$ and the energy range of the signal at all temperatures both indicate
that it is magnetic in origin. We observe this signal down to the lowest energies we were able to resolve in our experimental set-up on SEQUOIA, $\sim4$\,meV, but are unable to determine if there is an associated elastic diffuse signal. This signal was previously observed
at $(1\,0\,1)$ in the triple-axis-spectroscopy work of Radhakrishna and 
Cable~\cite{radhakrishna_inelastic-neutron-scattering_1996},
but we have now been able to map its full $Q$ and $E$ dependence
and confirm the signal's magnetic origin, in addition to observing
it at $(0\,0\,1)$. 

\begin{figure}
\includegraphics[clip,width=0.6\columnwidth]{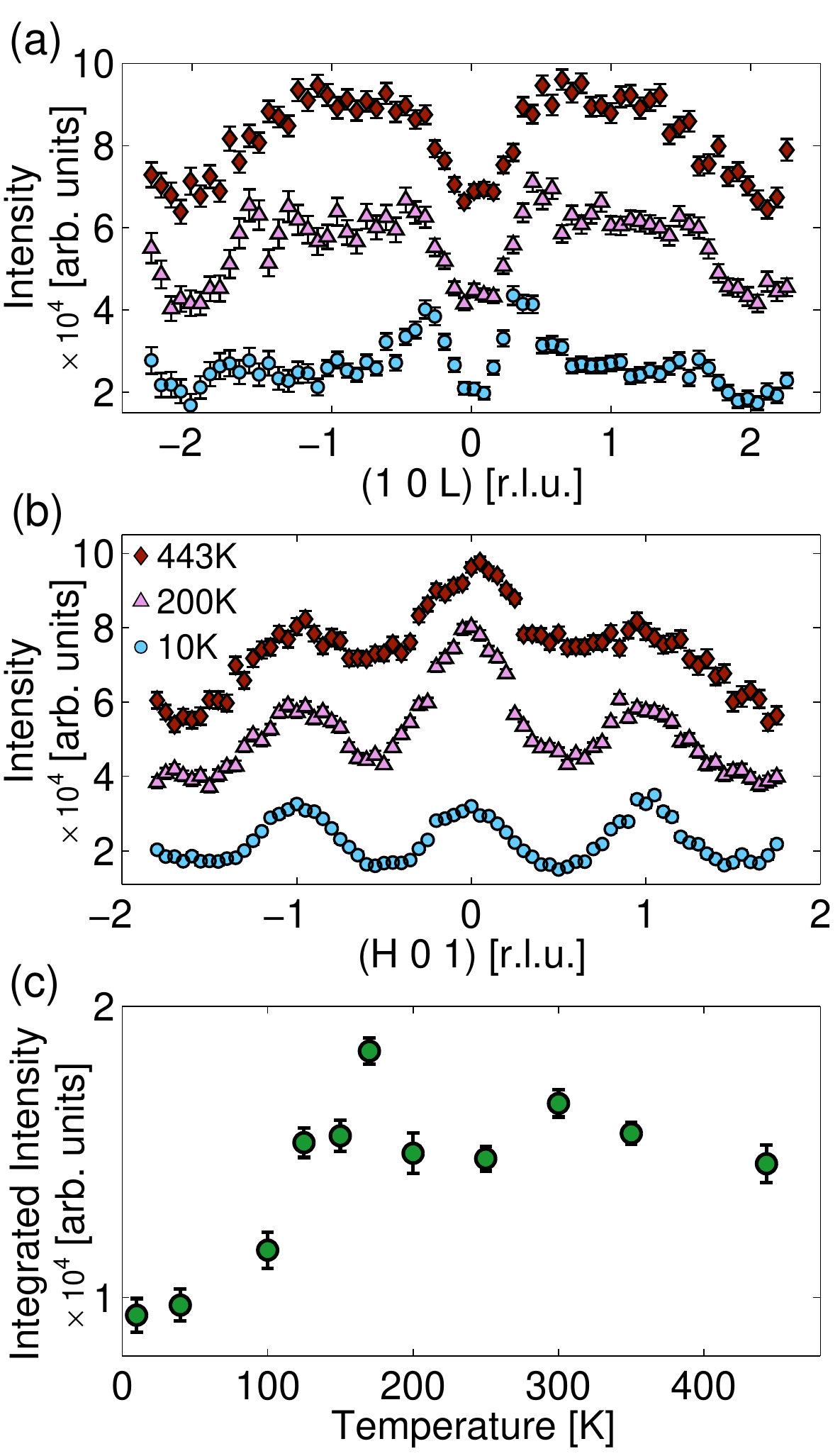}
\protect\caption{\label{fig:Blob_fits}(Color online) (a) and (b) Constant-energy cuts averaged over
20--25\,meV from $E_{\mathrm{i}}=150\,$meV SEQUOIA data measured
at 10\,K (circles), 200\,K (triangles) and 443\,K (diamonds).
Cuts in (a) were averaged over $\pm0.1\,$r.l.u. in perpendicular
$\bm{Q}$-space directions, and successive cuts were offset $2\times10^{-4}$
for clarity. Cuts in (b) were averaged over $\pm0.2\,$r.l.u. in perpendicular
$\bm{Q}$-space directions. (c) shows the temperature dependence of
the signal centered on $(0\,0\,1)$, determined from $(\mathrm{H}\,0\,1)$
cuts as described in the text. The plotted integrated intensity has
been corrected for the Bose population factor. }
\end{figure}

Further investigation of the scattering centered on $(0\,0\,1)$ is presented in
Fig.~\ref{fig:Blob_fits}. Cuts taken along $(1\,0\,\mathrm{{L}})$
and $(\mathrm{{H}}\,0\,1)$ directions through datasets collected
at 10\,K, 350\,K and 443\,K are shown in Fig.~\ref{fig:Blob_fits}(a) 
and~(b). As the $(\mathrm{{H}}\,0\,1)$ direction is the zone
boundary for the spin wave signal (see Fig.~\ref{fig:Const_E_slices}(c)) the cuts in Fig.~\ref{fig:Blob_fits}(b) and the color map in Fig.~\ref{fig:Dispersions}(c) essentially show
the broad signal in isolation. The $(1\,0\,\mathrm{{L}})$ direction, however,
passes through the zone center for both the spin wave and the broad
signal, therefore both are observed in  Figs.~\ref{fig:Blob_fits}(a) and~\ref{fig:Dispersions}(d). We show the temperature
dependence of the $(0\,0\,1)$ signal in Fig.~\ref{fig:Blob_fits}(c). 
The integrated intensity was determined from cuts made along
the $(\mathrm{{H}}\,0\,1)$ direction through the limited-angular-range
data measured at $E_{\mathrm{i}}=150\,$meV for all temperatures.
The data were averaged over 10--20\,meV and folded along H and L
to improve statistics. Two Gaussian functions on a flat background
were fit to the data, with widths constrained to be equal, but amplitudes
varying independently, and centers fixed at H=0 and H=1.
The resulting area of the Gaussian centered at H=0 was corrected for
the Bose population factor, $[1-\mathrm{exp}(-E/k_{B}T)]^{-1}$, at
each temperature and the result gives the integrated intensity plotted
in Fig.~\ref{fig:Blob_fits}(c).

These results show that the broad, $\bm{Q}=(0\,0\,1)$-centered scattering is influenced by \TSR{}, Fig.~\ref{fig:Blob_fits}(c). 
There is a dramatic increase in intensity of the signal on warming past \TSR{},
which is expected for scattering from a transverse magnetic fluctuation
as the spins reorient from $ab$-plane to $c$-axis alignment, because
the magnetic cross section depends only on the component of the magnetic
moment perpendicular to $\bm{{Q}}$. This is further supported by the observation that the (0\,0\,1) scattering is more sensitive to \TSR{} than the scattering at ($\pm$1\,0\,1), compare the cuts in Fig.~\ref{fig:Blob_fits}(b). These observations are a further confirmation
of the magnetic origin of the signal, and indicate coupling between this scattering and the ferromagnetic state of the system.

\subsection*{Spin-wave modelling}

We attempted to model the inelastic magnetic response of the system 
using linear spin-wave theory. Isotropic exchange interactions including up to sixth nearest neighbors were the minimum 
required to model the spin-wave frequencies. The six nearest neighbor exchange parameters, $J$s, are illustrated in Fig.~\ref{fig:SpinWaveCalculations}(c).  Single-ion anisotropy was not included because there was 
no evidence for a gap in the spin-wave spectrum down to at least 0.5\,meV (<1\,\% of the bandwidth), see Fig.~\ref{fig:CTAX}. The inelastic neutron cross section for undamped spin waves was calculated using the 1/S formalism outlined in Ref.~\onlinecite{Haraldsen2009} and appendix A of Ref.~\onlinecite{Fishman2013}. For comparison with experimental intensities, the effects of the magnetic form factor 
and an approximation for the instrumental resolution were included in the calculation. We used the magnetic form factor for Mn$^{2+}$ from 
Ref.~\onlinecite{dianoux_neutron_2003}. The resolution function was approximated as a Gaussian in energy with a
full width at half-maximum of 7.5 meV and the results averaged over the same volume of $\boldsymbol{Q}$ as the
experimental data (Fig.~\ref{fig:Dispersions}) and averaged over six domains. To determine the exchange parameters, the model was fit to the dispersions along (1\,0\,0), (0\,0\,1) and (1\,1\,0) directions, extracted from the experimental data via a series of constant-energy cuts and fits like those in Fig.~\ref{fig:Dispersions}(d). We determined the energy at the (H\,0\,2) zone boundary, 70(4)\,meV, by fitting a Gaussian to a constant-$\boldsymbol{Q}$ cut at (1.5\,0\,2) from which a similar cut at (2\,0\,2) had been subtracted as a background. 

The results of this modeling are shown in Fig.~\ref{fig:SpinWaveCalculations}(a) and (b), which show reasonable correspondence to the spin-wave data presented in Fig.~\ref{fig:Dispersions}(a) and (b). However, the bandwidth of the dispersion does not appear to be well reproduced along the H direction, see Fig..~\ref{fig:SpinWaveCalculations}(a).  Only $J_6$ can be uniquely determined from our results. Instead, the linear combinations   
\begin{align*}
\alpha &= J_1+ 6J_3 \\
\beta &= J_4 + 6J_5 \\ 
\gamma &= J_2 + 2J_3 +2J_5 \\
\end{align*}
were fit. The values best fitting these measurements are $\alpha S$= 20.0 meV, 
$\beta S$ = \mbox{-0.8 meV}, $\gamma S$= 4.8 meV,  and $J_6 S$ = 2.9 meV.  Fig.~\ref{fig:SpinWaveCalculations}
shows the calculated frequencies and intensities along two of the measured directions, for comparison to Fig.~\ref{fig:Dispersions}(a) and (b).  Previously Refs.~\cite{yamaguchi_magnons_1992} and~\cite{radhakrishna_inelastic-neutron-scattering_1996} reported triple-axis spectroscopy measurements on \MnSb{}  which covered limited ranges of $\bm{Q}$ and $E$, not reaching the zone boundary. In Ref~\cite{yamaguchi_magnons_1992} they fit a Heisenberg model up to fourth nearest neighbors (i.e. $J_5=J_6 = 0$ for comparison to our model) to the low energy data measured on a Mn$_{1.05}$Sb crystal. They found $\alpha S = 8.75$\,meV, 
$\beta S = 2.4$\,meV and $\gamma S= 5.4$\,meV. These results extrapolate to a bandwidth of 130\,meV along (H\,0\,2), which is much larger than the 70(4)\,meV we observe, see Fig.~\ref{fig:Dispersions}.

\section*{Discussion}

\subsection*{Magnetic Structure}

The results of our neutron diffraction experiment on \MnSbx{} are
well described by the magnetic structure shown in Fig.~\ref{fig:Diffraction}(c), 
with ferromagnetic alignment of Mn1 ions, and Mn2 ions aligned antiparallel to Mn1. 
There appears to be no need to include
the model of an aspherical form factor for Mn1 to account for
magnetic scattering at $l=\mathrm{odd}$, $(h-k)\neq3n$ Bragg peak positions for our data
set. This is consistent with early neutron powder diffraction results
on \MnSb{}~\cite{bouwma_neutron_1971}, but in contrast to the interpretations
of Yamaguchi \etal{}~\cite{yamaguchi_polarized_1978} and Reimers
\etal{}~\cite{reimers_polarised_1983} from their polarized-neutron,
single-crystal experiments. 

The single crystals investigated in Refs.~\cite{yamaguchi_polarized_1978} and~\cite{reimers_polarised_1983} 
were of composition Mn$_{1.05}$Sb and Mn$_{1.09}$Sb respectively,
lower interstitial contents than our crystal, and in both cases they
excluded some low $Q$ data from their analysis. This may explain
the seeming discrepancy between their interpretation and our results.
In the former case, the scattering from Mn2 sites would have been weaker,
and therefore harder to interpret. The higher $Q$ data that they
included in their analysis is less sensitive to magnetic scattering
from the interstitial site, but more sensitive to the asymmetric form
factor they describe. In the case of our data, only model 1, including an antiparallel
moment on Mn2, can account for all peaks observed below \TC{}
across all $Q$, see Fig.~\ref{fig:Fullprof}(c)--(f). Our data
is less sensitive to any asymmetry in the Mn1 form factor because
it was not a polarized neutron measurement and the magnetic contribution at
high $Q$ is small. Our results do not preclude
some asphericity in the magnetic form factor of Mn1. It is likely,
however, that the asphericity found in the previous models is too
large, as some of the scattering intensity should have been accounted
for by the moment on Mn2. This helps to explain the discrepancies
between the Yamaguchi model~\cite{yamaguchi_polarized_1978} and results
from x-ray Compton profile analysis~\cite{nakamura_magnetic_1995},
and electronic structure calculations~\cite{coehoorn_electronic_1985-1},
as well as the failure of the model at low~$Q$~\cite{yamaguchi_magnetic_1976,yamaguchi_polarized_1978,reimers_polarised_1983}. 

The size of the magnetic moments
found from our FullProf refinement of model 1 are consistent with
magnetization measurements. The saturated moment per formula unit was found to be 
3.20$\,\mu_{\mathrm{B}}$ at 2\,K from magnetization. The results from HB-3A give a
total moment per formula unit of 3.2(1)$\,\mu_{\mathrm{B}}$ at 10\,K, taking into account
the Mn2 occupancy of $\delta=0.13(1)$, see Table~\ref{tab:Diffraction}. The determination of a
magnetic moment associated with the Mn2 site is further supported
by the results from our DFT calculation. Previous modeling of
\MnSb{} did not include an interstitial Mn, and those that attempted
to account for it merely adjusted the electron count in stoichiometric MnSb
models~\cite{coehoorn_electronic_1985,continenza_coordination_2001,katoh_electronic_1986,podloucky_electronic_1984,ravindran_magnetic_1999,sandratskii_energy_1981,vast_first-principles_1992,chen_properties_1978}. We find a sizable moment on Mn2 in the calculation, see Fig.~\ref{tom1}.

Given the presence of a magnetic moment on the interstitial site, in a simple local Heisenberg picture
we would expect the antiparallel alignment of the interstitial to enhance the overall ferromagnetic state. However, it has clearly
been observed that the presence of the interstitial acts to reduce \TC{}~\cite{chen_properties_1978,okita_crystal_1968,teramoto_existence_1968}.
Our DFT calculation shows that the interstitial Mn degrades the stability of the ferromagnetic state beyond the simple influence of changing lattice parameters (see Fig.~\ref{tom2}),  in agreement with the experimental observations. This indicates that a purely localized moment model may not be sufficient to describe the magnetic state in $\delta\neq0$ \MnSb{}. 

\subsection*{Magnetic Dynamics}

Our inelastic neutron scattering measurements of \MnSb{} serve to highlight the influence the interstitial
Mn has on the magnetic state. The spin-wave-type excitations appear
to be less well defined than in the non-interstitial containing sister
compound MnBi (see Fig.~\ref{fig:Const_E_slices}), and are modified
such that a simple Heisenberg model does not describe the observed
dispersion. For a local moment system, linear spin-wave theory is
expected to capture the form of the dispersion by inclusion of a small number of 
nearest neighbor exchange parameters. Although
the observed dispersion (e.g. Fig.~\ref{fig:Dispersions}~(a))
appears to be associated with the ferromagnetic wavevector, and a similar feature is
observed in interstitial-free MnBi (Fig.~\ref{fig:Const_E_slices}(b)~\cite{williams_unpub}), we have shown that the measurement is not
fully described by the Heisenberg model for localized Mn1 spins, Fig. \ref{fig:SpinWaveCalculations}. 

For the spin wave calculation along (H\,0\,2), attempting to match both the bandwidth
and the gradient of the dispersion at lower energies requires a large $J_6$ interaction (2.9\,meV), which probably reflects the inadequacy of the model.
The shape at the zone boundary remains quite different, resulting in the 70(4)\,meV bandwidth not being reproduced, see Fig.~\ref{fig:SpinWaveCalculations}(a). The calculation does not produce a signal emanating from (0\,0\,1) as is observed in the data, see Fig.~\ref{fig:Dispersions}(c). As (H\,0\,1) is a zone boundary for the Mn1 magnetic lattice, as illustrated in Fig.~\ref{fig:Const_E_slices}(c), the spin-wave intensity at (0\,0\,1) is concentrated around the zone boundary energy $E=85\,meV$ (see Fig.~\ref{fig:SpinWaveCalculations}(b)). Our crude attempts to extend the 
model to include an interstitial Mn2 moment failed to better reproduce the spin wave data, 
nor did they produce a signal like the observed scattering along $(0\,0\,1)$. These attempts are limited at the present time by the absence of general theoretical and computational tools to include the effects of a disordered component in a system
on the inelastic magnetic response. Including a small moment at one interstitial site in our
model calculation artificially introduces a symmetry element
which is not present in the real system. Extending the calculation
to include thousands of unit cells, which can then be randomly assigned interstitials,
is currently difficult to implement and extremely computationally intensive.

In addition to the spin wave, we identified a second magnetic signal 
in the inelastic spectrum which is broad in $Q$, and centered on an unexpected
position in reciprocal space $\bm{{Q}}=(0\,0\,1)$ (see Figs.~\ref{fig:Const_E_slices},
\ref{fig:Dispersions} and~\ref{fig:Blob_fits}). Like the spin-wave scattering, this signal responds to \TSR{}, but unlike the spin-wave scattering,  no similar signal is observed in interstitial-free MnBi. $(0\,0\,1)$ is
not an allowed nuclear or magnetic reflection (for either magnetic
structure model discussed above). The presence of a $(0\,0\,1)$ Bragg
peak would require the correspondence between the upper and lower
halves of the unit cell shown in Fig.~\ref{fig:Diffraction}(c) to
be broken, for example, if Mn1 was antiferromagnetically aligned
along $c$, or if Mn2 was only present in the lower half of
the cell. We do not observe a Bragg peak at $(0\,0\,1)$
in any of the neutron data sets collected, implying that no such long-range
order is present. However, the periodicity of the signal in reciprocal space implies that underlying long-range correlations are present.

Our results show that the dynamic correlations observed at $(0\,0\,1)$ are not
associated with a long-range order in the system. They could,
however, be associated with a short-range structural or magnetic order
in the system. This proposition may be supported by the broad appearance
of the signal in reciprocal space (see Fig.~\ref{fig:Const_E_slices}),
which suggests a short correlation length in real space. We made a
rough estimate for the correlation length by fitting Gaussians to
cuts similar to those in Fig.~\ref{fig:Blob_fits}(a) and~(b), integrated
over 10--15\,meV, from the $E_{\mathrm{i}}=60\,$meV, $T=10\,$K
data. We find correlation lengths of \textasciitilde{}16$\,\mathrm{\AA}$
in the $a$-$b$ plane, and in the range 10--18$\,\mathrm{\AA}$ along
$c$ (where it is harder to define, due to the overlap of the signal
with the spin-wave-type scattering), i.e. \textasciitilde{}4 unit
cells in plane, and \textasciitilde{}3 units cells along $c$. We
were unable to determine whether there is an elastic, diffuse scattering
signal at the $(0\,0\,1)$ position due to the neutron instrumentation
used in this investigation. We do observe the signal down to the lowest
energies that were probed on SEQUOIA, $\sim4$\,meV. A short-range order could be present in
the system if the interstitial Mn ions are not randomly distributed, but
instead are structurally ordered over a few unit cells. An observation of elastic, diffuse scattering at (0 0 1) would establish that short-range order is present. 
Alternatively, the presence of interstitials could modify the interactions of neighboring
main-site Mn1 ions, with the associated magnetic excitations unable to
propagate over large length scales due to disruption from a random
distribution of magnetic Mn2 ions. 
Further studies, including diffuse neutron scattering
experiments, are highly desirable to investigate the potential short-range
order in this system. 

The spectral weight of the $(0\,0\,1)$ scattering, i.e. the total
intensity of the signal integrated over $\bm{{Q}}E$ space, can in
principal be used to indicate the strength of the magnetic excitations
and therefore the size of the fluctuating magnetic moment, which could
give an indication of the origin of the scattering. Unfortunately,
due to the overlap of the two magnetic excitations in $\bm{{Q}}E$
space, we cannot quantitatively compare the spectral weights
of the two signals. Qualitatively, however, by inspecting a series
of slices and cuts through the data, such as those shown in Figs.~\ref{fig:Const_E_slices}, 
\ref{fig:Dispersions} and~\ref{fig:Blob_fits},
we see that the strength of the signal emanating from $(0\,0\,1)$
is comparable to the spin-wave signal. This implies that
the (0\,0\,1) signal is associated with a relatively large magnetic moment per
formula unit in the sample, of the order of the Mn1 moment size.
This makes it unlikely to be purely the result of the Mn2
moments behaving independently of Mn1. Identification of the
origin of this signal would be extremely significant for understanding
the magnetic state in \MnSb{} and determining the possibilities of tuning the state with
interstitial ions of Mn or other 3$d$ transition metals.

\section*{Conclusions}

We have shown that a magnetic moment on the interstitial Mn in \MnSb{}
is required to describe the neutron diffraction data from a single
crystal of \MnSbx{}. This magnetic moment is aligned antiparallel
to the main site Mn magnetic moment. We performed DFT calculations which find that the interstitial Mn is magnetic, consistent with the diffraction data.
The DFT results are also consistent with previous experimental evidence that the interstitial reduces \TC{} and $c$ lattice parameter, and increases $a$  in \MnSb{}. 
We find that the presence of the magnetic interstitial Mn also has a substantial effect on the magnetic excitation spectrum
of \MnSb{}, measured by inelastic neutron scattering. It results in the appearance of an intense, broad signal in $Q$-space, that is likely associated with short-range order. While a Heisenberg Hamiltonian model calculation captures the essential $Q$-space dependence of the spin-wave type signal observed, the additional broad signal cannot be explained by straightforward  attempts to include an interstitial ion in this model.

\section*{Acknowledgements}
We thank J.~Q.~Yan, D.~Mandrus, M.~A.~McGuire and B.~C.~Sales for useful discussions.
The research at ORNL's Spallation Neutron Source and High Flux Isotope Reactor was supported by the Scientific User Facilities Division, Office of Basic Energy Sciences, U.S. Department of Energy (DOE). 
A.F.M. and R.F. were supported by the U.S. DOE, Office of Science, Basic Energy Sciences, Materials Sciences and Engineering Division. 
T.B. and T.J.W. are supported as Wigner Fellows at ORNL. Work by TB was performed at the Center for Nanophase Materials Sciences, a DOE Office of Science user facility. This research used resources of the National Energy Research Scientific Computing Center, a DOE Office of Science User Facility supported by the Office of Science of the U.S. DOE under Contract No. DE-AC02-05CH11231. S.E.H. acknowledges support by the Laboratory's Director's fund, ORNL.

This manuscript has been authored by UT-Battelle, LLC under Contract No. DE-AC05-00OR22725 with the U.S. Department of Energy.  The United States Government retains and the publisher, by accepting the article for publication, acknowledges that the United States Government retains a non-exclusive, paid-up, irrevocable, world-wide license to publish or reproduce the published form of this manuscript, or allow others to do so, for United States Government purposes.  The Department of Energy will provide public access to these results of federally sponsored research in accordance with the DOE Public Access Plan (http://energy.gov/downloads/doe-public-access-plan).

\bibliographystyle{apsrev4-1}

\bibliography{MnSb_bib}

\begin{thebibliography}{50}
\bibitem [{U8e()}]{U8ev}%
  \BibitemOpen
  \href@noop {} {}\bibinfo {note} {For $U=8$\,eV we find that the energy
  difference between the FM and the AFM-c configuration reduces from 212 to
  156\,meV and the energy difference between the between the FM and the AFM-a
  configuration lowers from 219 to 185\,meV.}\BibitemShut {Stop}%
\end{thebibliography}

\end{document}